\documentclass[twocolumn]{aastex631}

\newcommand{\kms}{{\rm km\,s}\ensuremath{^{-1}}}
\newcommand{\radioqso}{P352--15}

\newcommand{\cii}{[\ion{C}{2}]}

\submitjournal{ApJ Letters}

\shorttitle{Radio jet--gas alignment in a z\,$\sim$\,6 quasar}
\shortauthors{Walter et al.}

\begin{document}

\title{Kiloparsec--Scale Alignment of a Radio Jet with Cool Gas and Dust in a z$\sim$6 Quasar}

\correspondingauthor{Fabian Walter}
\email{walter@mpia.de}

\author[0000-0003-4793-7880]{Fabian Walter}
\affiliation{Max Planck Institut f\"ur Astronomie, K\"onigstuhl 17, D-69117 Heidelberg, Germany}
\affiliation{California Institute of Technology, Pasadena, CA 91125, USA}
\affiliation{National Radio Astronomy Observatory, Pete V. Domenici Array Science Center, P.O. Box O, Socorro, NM 87801, USA}

\author[0000-0002-2931-7824]{Eduardo Ba\~nados}
\affiliation{Max Planck Institut f\"ur Astronomie, K\"onigstuhl 17, D-69117 Heidelberg, Germany}

\author[0000-0001-6647-3861]{Chris Carilli}
\affiliation{National Radio Astronomy Observatory, Pete V. Domenici Array Science Center, P.O. Box O, Socorro, NM 87801, USA}

\author[0000-0002-9838-8191]{Marcel Neeleman}
\affiliation{National Radio Astronomy Observatory, 520 Edgemont Road, Charlottesville, VA, 22903, USA}

\author[0000-0002-7898-7664]{Thomas Connor}
\affiliation{Center for Astrophysics $\vert$\ Harvard\ \&\ Smithsonian, 60 Garden St., Cambridge, MA 02138, USA}

\author[0000-0002-2662-8803]{Roberto Decarli}
\affiliation{INAF—Osservatorio di Astrofisica e Scienza dello Spazio, via Gobetti 93/3, I-40129, Bologna, Italy}

\author[0000-0002-6822-2254]{Emanuele Paulo Farina}
\affiliation{GEMINI Observatory Northern Operations Center, 670 N.\ A'ohoku Place - Hilo, Hawaii, 96720, USA}

\author[0000-0002-7220-397X]{Yana Khusanova}
\affiliation{Max Planck Institut f\"ur Astronomie, K\"onigstuhl 17, D-69117 Heidelberg, Germany}

\author[0000-0002-5941-5214]{Chiara Mazzucchelli}
\affiliation{Instituto de Estudios Astrof\'{\i}sicos, Facultad de Ingenier\'{\i}a y Ciencias, Universidad Diego Portales, Avenida Ejercito Libertador 441, Santiago, Chile}

\author[0000-0001-5492-4522]{Romain Meyer}
\affil{Max Planck Institut f\"ur Astronomie, K\"onigstuhl 17, D-69117 Heidelberg, Germany}
\affiliation{Department of Astronomy, University of Geneva, Chemin Pegasi 51, 1290 Versoix, Switzerland}

\author[0000-0003-3168-5922]{Emmanuel Momjian}
\affiliation{National Radio Astronomy Observatory, Pete V. Domenici Array Science Center, P.O. Box O, Socorro, NM 87801, USA}

\author[0000-0003-4996-9069]{Hans--Walter Rix}
\affiliation{Max Planck Institut f\"ur Astronomie, K\"onigstuhl 17, D-69117 Heidelberg, Germany}

\author[0000-0003-2349-9310]{Sof\'ia Rojas-Ruiz}
\affiliation{Department of Physics and Astronomy, University of California, Los Angeles, 430 Portola Plaza, Los Angeles, CA 90095, USA}

\author[0000-0001-9024-8322]{Bram Venemans}
\affiliation{Leiden Observatory, Leiden University, PO Box 9513, 2300 RA Leiden, The Netherlands}

\begin{abstract}
We present high--angular resolution ($0.068^{\prime\prime}$, $\sim$400\,pc) ALMA imaging of the \cii\ line and dust continuum emission of PSO~J352.4034--15.3373, a radio--loud quasar at $z$\,=\,5.83. The observations reveal a remarkably close match between the orientation of the \cii\ and thermal dust emission mapped by ALMA, and radio synchrotron emission of a radio jet previously mapped by the VLBA. This narrow alignment extends over $\sim$4\,kpc, reminiscent of the well--studied `alignment effect' in lower--redshift radio galaxies. The \cii\ kinematics show a linear increase in velocity with galactocentric radii up to $\sim200$\,km\,s$^{-1}$ at $r$\,=\,2\,kpc, consistent with bulk motions within the galaxy potential, and not relativistic jet motions. The kinematics and respective morphologies are consistent with a picture in which the relativistic jet injects energy into the interstellar medium (potentially leading to subsequent star formation), giving rise to the observed alignment and significant ($\gtrsim 100$\,km\,s$^{-1}$) \cii\ velocity dispersion within the host galaxy on kiloparsec scales.  Indeed, the astonishingly close alignment and narrow linearity of the radio jet with the \cii\ and dust emission are hard to conceive without some fundamental relationship between the two.
 \end{abstract}


\section{Introduction}
\label{sec:intro}
Ever since their first detection more than 20 years ago \citep{Fan01}, the existence of quasars at Cosmic Dawn ($z$\,$\gtrsim$\,6; i.e., when the Universe was $<$\,1\,Gyr old) has challenged our understanding of early supermassive black hole and galaxy formation  \citep[e.g., review by][]{Fan23}. Their high rest-frame-UV brightness and broad line region (BLR) properties provide clear evidence that the centers of these quasars are powered by accreting supermassive black holes (SMBHs) that often exceed masses of 10$^{9}$\,M$_{\odot}$. These SMBHs thus must have accreted copious amounts of mass throughout their short lifetimes. To explain this rapid growth, theories invariably predict that quasars emerge in overdense environments through rapid gas accretion and mergers. These processes also build up the host galaxies, but theories disagree if the growth of the black holes outpaces the growth of the galaxies or vice-versa \citep[e.g.,][]{Volonteri21, Habouzit21}

One of the early puzzling findings from the pre--ALMA era was the detection of large amounts of cold dust and molecular gas in the host galaxy of the $z\!=\!6.4$ quasar J\,1148+5251 \citep[exceeding gas masses of 10$^{10}\,$M$_\odot$, ][]{Walter03, Walter04, Bertoldi03, Maiolino05}, demonstrating the presence of highly enriched gas on kiloparsec scales within the first Gyr of the Universe. Today, out of the $\sim$\,300 quasars that are known at redshifts $z\!\gtrsim\!6$, about 100 have been observed with ALMA, and most are detected through the redshifted 158\,$\mu$m line of ionized carbon (\cii) and the underlying thermal dust continuum \citep[e.g.,][]{Decarli18, Venemans18, Fan23}. This interstellar medium (traced through dust continuum, \cii\ and CO emission lines) is found to be in a cold phase \citep[$\sim$\,50\,K, e.g.,][]{Leipski14} and is thought to be heated by star formation in the quasar hosts. With the advent of JWST, star formation and stellar mass tracers are now also being routinely detected in this high--redshift quasar population \citep[e.g., ][]{Ding23, Decarli24, Onoue24}.

So far, high--spatial--resolution (few--100\,pc) ALMA imaging of the dust and \cii\ in $z\,\gtrsim\,6$ quasar host galaxies have found little evidence for interactions or feedback between the central SMBH and the host galaxy \citep{Venemans19, Walter22, Meyer23, Neeleman23, Meyer2025}. This is surprising, as the active galactic nucleus (AGN) or quasar phase of a galaxy, i.e., when the central SMBH is actively accreting material and releasing energy into the surrounding medium, is thought to play an important role in shaping its host galaxy, thus affecting the process of galaxy evolution \citep[e.g.,][]{Fabian12, Wylezalek18}. For example, relativistic jets powered by SMBHs are thought to affect their environment on galactic (kpc or even 10s of kpc) scales (the so-called kinetic or radio-mode feedback). This kinetic feedback is often associated with low-power AGN, but this mode can also be significant on sources with high accretion rates, such as quasars \citep[e.g.][]{Russell13, Nesvadba17}. 

In this Letter, we present high-resolution ALMA observations to map the dust and \cii\ gas of the $z\!=\!5.83$ quasar PSO J352.4034–15.3373 \citep[hereafter \radioqso,][]{banados2018c}. This quasar was originally selected through standard color-selection techniques and later identified to be one of the few radio--loud quasars known at the time at $z\!\sim\!6$. 
Very Large Baseline Array (VLBA) observations at high resolution ($\sim$0.017$\arcsec$, $\sim$100\,pc) revealed that \radioqso\ hosts the largest jet observed in the Universe's first billion years \citep{Momjian18}, with X-ray observations showing that the jet may extend as much as 50 kpc \citep{Connor21}. Follow--up low (kpc--scale) resolution ALMA \cii\ and dust continuum mapping of \radioqso\ revealed typical gas and dust masses as compared to the $z\!\sim\!6$ quasar population: indeed, \radioqso\ broadly follows the L$_{\rm FIR}$ -- L$_{\rm [CII]}$ distribution found in the $z$\,$\sim$\,6--7 quasar distribution. 
The FIR--radio spectral energy distribution implies that the ALMA continuum emission is not significantly affected (at most at a 10\% level) by the synchrotron emission seen in the radio regime \citep{Rojas-Ruiz21}.

We use a flat cosmology with $H_0 = 70 \,\mbox{km\,s}^{-1}$\,Mpc$^{-1}$, $\Omega_M = 0.3$, and $\Omega_\Lambda = 0.7$.  At $z=5.831$, the Universe is 950 Myr old, and one proper kpc corresponds to $0\farcs 172$ ($1\arcsec$ equals 5.81\,kpc).

\begin{figure*}
\includegraphics[width=\textwidth]{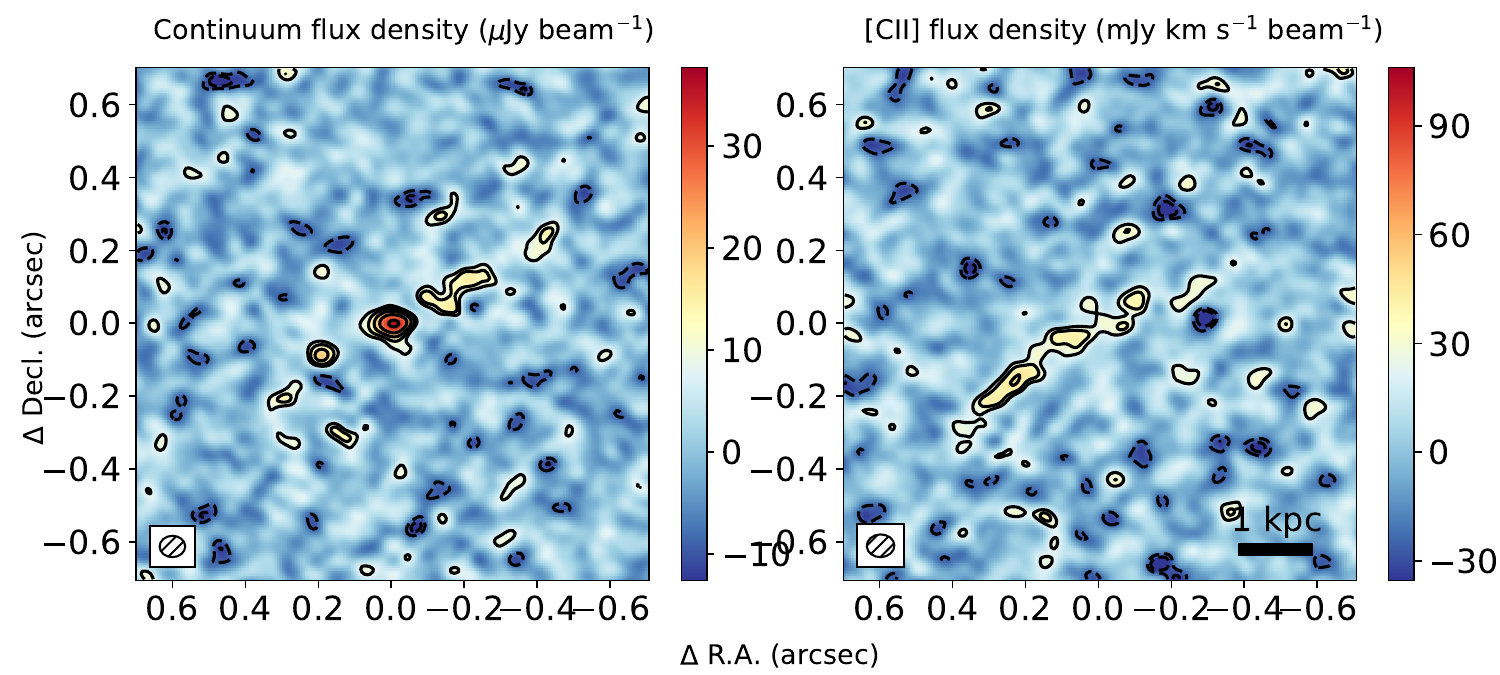}
\caption{ALMA 400\,pc resolution imaging of the dust continuum emission at 285 GHz (\textit{left}) and the \cii\ emission averaged over the central 600\,km\,s$^{-1}$ (\textit{right}). Contours start at $\pm2\sigma$ and increase/decrease in powers of $\sqrt{2}$. For the continuum, 1\,$\sigma$ corresponds to 4.1$\mu$Jy\,beam$^{-1}$, for the \cii\ line, 1\,$\sigma$\,=11.8\,mJy\,km\,s$^{-1}$\,beam$^{-1}$. Coordinates are given relative to the brightest source seen in the VLBA observations discussed in Sec.~\ref{sec:vlba} at R.A.(ICRS)=$23^{\rm h}29^{\rm m}36 \fs 8316$, Dec.(ICRS)$=-15^{\circ}20^{\prime}14 \farcs 498$ \citep{Momjian18}. The beamsize is indicated in the lower left corner of both panels, and a 1\,kpc scale is shown in the bottom right of the \cii\ map.}
\label{fig:alma}
\end{figure*}

\section{Observations}
\label{sec:observations}
\radioqso\ was observed in ALMA Band 7 \citep{Mahieu12} to map its (rest--frame) 158\,$\mu$m \cii\ emission line and underlying continuum emission. The observations were carried out during two different ALMA Cycles (cycles 6 and 9) with similar antenna configurations (5$^{\rm th}$ [80$^{\rm th}$] percentiles of 336\,m [3626\,m] and 315\,m [3282\,m], respectively). The first set of observations was conducted in Cycle 6 on UT 2019, July 31 (ALMA PID  2018.1.00656.S) for a total duration of 1.7 hours (49 minutes on source). The second set of observations was conducted in Cycle 9 between UT 2023 June 22 and June 2023 June 30 (ALMA PID 2022.1.01587.S) with a total time of 29.8 hours (12.9 hours on source). The blazar J0006$-$0623 was used as a flux density scale and bandpass calibrator, and the blazar J2331$-$1556 was used as a phase calibrator. One of the 1.875\,GHz spectral windows was centered on the expected frequency of the redshifted \cii\ emission (277.8 GHz) whereas the remaining 1.875 GHz spectral windows were set up to detect continuum emission at sky frequencies of 279.8, 289.9 and 291.9 GHz.

The data from 2018.1.00656.S and 2022.1.01587.S  were processed using the Cycle-specific ALMA pipelines \citep{hunter2023}, which are part of the Common Astronomy Software Application (CASA) package \citep[V.5.6.1-8 and V6.2.1-7 for Cycle 6 and Cycle 9, respectively;][]{casa2022, mcmullin2007}. The two data sets were then combined using the task \textit{concat} in the most recent version of CASA at the time of analysis (V6.5.2-26). Further imaging and processing was done in this version of CASA as well.

A continuum image was created using the CASA task \textit{tclean} by considering only the line--free channels within the data. This resulted in a total bandwidth of 6.9 GHz after removing 0.6 GHz around the \cii\ emission line. For all of the imaging, we used Briggs weighting with a robust factor of 0.5. To reduce imaging artifacts caused by the sparser sampling at longer baselines, we also applied a $uv$-taper corresponding to a Gaussian with a standard deviation in the image plane of 0\farcs04. This results in a continuum image with a root mean square (rms) noise level of 4.1~$\mu$Jy~beam$^{-1}$ and a synthesized beam size of 0\farcs076 $\times$ 0\farcs062 (corresponding to 0.44\,kpc $\times$ 0.36\,kpc at the redshift of \radioqso). 

To create the spectral cube, we first subtracted the underlying dust continuum using the task \textit{uvcontsub} by fitting a zeroth order polynomial to the line-free channels in the spectral window containing the redshifted \cii\ emission line. The spectral cube was then created within \textit{tclean} with channel widths of 31.25\,MHz ($\sim$\,34\,km\,s$^{-1}$) resulting in an average rms noise of 62~$\mu$Jy~beam$^{-1}$ per channel and a synthesized beam of 0\farcs076 $\times$ 0\farcs063. We also generated an integrated \cii\ map by imaging a single 600~\kms\ channel centered on the \cii\ emission line (i.e., centered at 278.2976~GHz). This velocity range covers the full \cii\ emission 
of \radioqso\ \citep[$\Delta$V$_{\rm FWHM}\,=435\pm 82$\,km\,s$^{-1}$,][]{Rojas-Ruiz21}. The rms noise of this \cii\ map is 19.6~$\mu$Jy~beam$^{-1}$ or 11.8\,mJy~\kms~beam$^{-1}$ and has the same synthesized beam as the spectral cube. We note that our imaging data have a maximum recoverable scale of 0\farcs6, implying that our data is not sensitive to emission on larger scales. 

\begin{figure*}
\begin{center}
\includegraphics[width=0.75\textwidth]{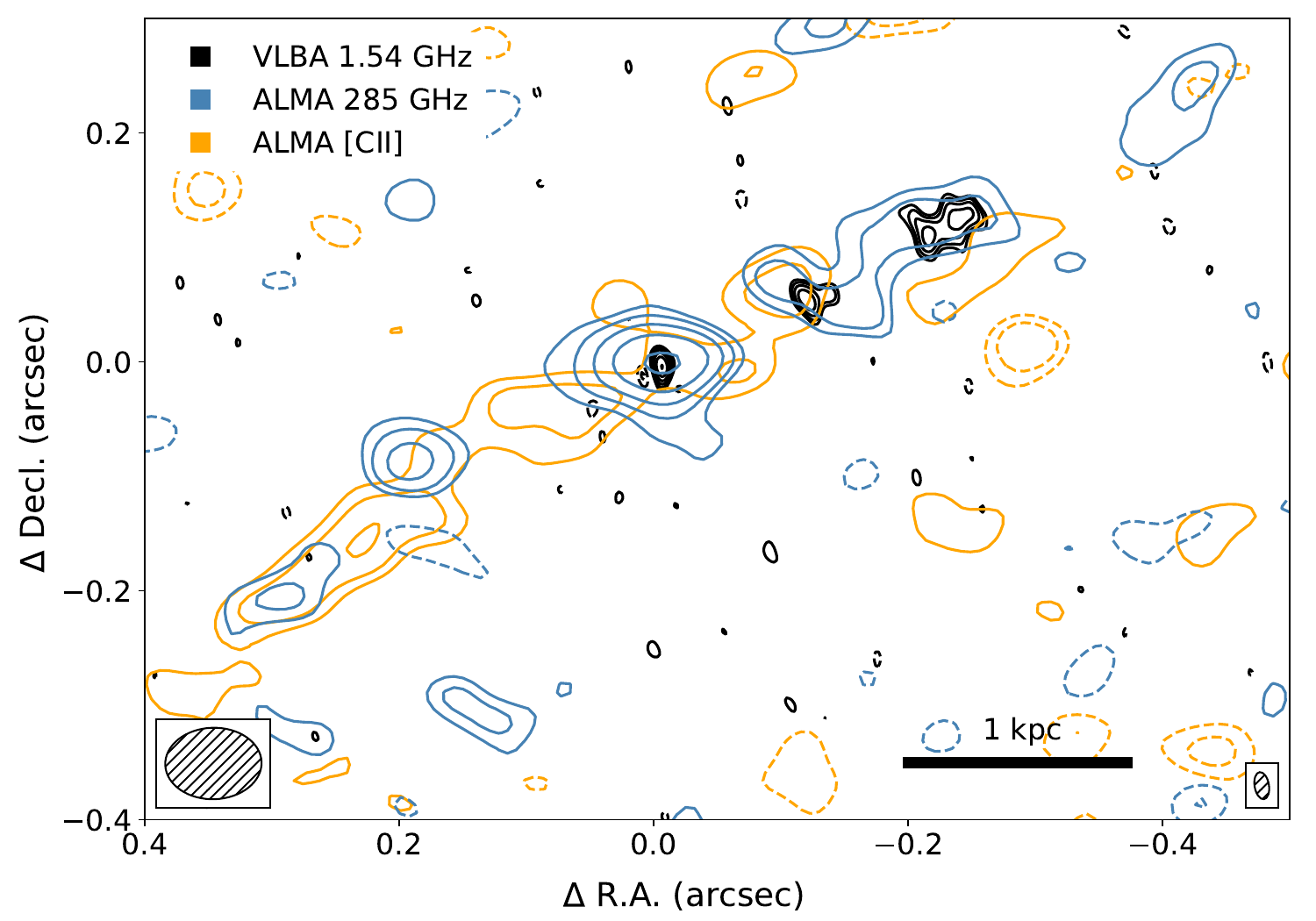}
\end{center}
\caption{
Superposition of the ALMA dust continuum (blue contours), ALMA \cii\ emission (orange contours), and the VLBA 1.54\,GHz radio continuum image (black contours). The dust continuum emission at 285\,GHz and \cii\ emission have the same contour levels as shown in  Fig.~\ref{fig:alma}. The VLBA image presented in \cite{Momjian18} is shown in black contours starting at $\pm$3$\sigma$ (1$\sigma$ = 67\,$\mu$Jy\,beam$^{-1}$) and increase/decrease by powers of $\sqrt{2}$. The ALMA (VLBA) beamsize is indicated in the lower left (right) corner. The origin in this figure, and the kpc--scale bar is as in Fig.~\ref{fig:alma}. As discussed in Sec.~\ref{sec:vlba} we have shifted the ALMA data slightly ($<\,25$\,mas) in order to center the ALMA dust continuum peak on the peak of the VLBA image.}
\label{fig:vlbi}
\end{figure*}

\section{Results} \label{sec:results}

\subsection{ALMA continuum emission}
\label{sec:cont}
The left panel of Fig.~\ref{fig:alma} shows the ALMA continuum map. With an rms noise of 4.1\,$\mu$Jy\,beam$^{-1}$, individual emission peaks with a significance of 3$\sigma$ and higher are located along a linear structure over 0\farcs93 (5.4 kpc). The peak emission is 37$\pm$4$\,\mu$Jy\,beam$^{-1}$, corresponding to 9$\sigma$. Summing up all emission along the linear feature gives a total flux density of 0.17\,mJy, or about 50\,\% of the flux reported in the lower--resolution (arcsec--scale) imaging presented by \cite{Rojas-Ruiz21}\footnote{We note that we recover the same total flux in our data when tapering the emission to a 0.3$"$ beam.}.
This implies that about half of the continuum emission is not recovered in our highest--resolution observations and is distributed on larger spatial scales, presumably in the host galaxy. It should be stressed that no other of the half-dozen $z\,\sim\,6$ quasars observed at similar sub--kpc spatial resolution show such a linear clumpy dust continuum structure \citep[e.g.,][]{Venemans19, Walter22, Meyer23, Neeleman23, Meyer2025}.

Averaged over the synthesized beam of $\Theta$=0.065$\arcsec$, $\lambda$=0.107\,cm (corresponding to 278.25\,GHz), and using equation 2.33 in \cite{Condon2016} and correcting for redshift, this central peak corresponds to a beam--averaged surface brightness temperature contrast of T$_B$=1.1\,K above the extended Cosmic Microwave Background (CMB), and the other peaks along the linear structure consequently have even lower surface brightness temperatures. This low temperature implies that the emission is beam-diluted thermal dust emission, and not due to non--thermal synchrotron emission. The latter would have an intrinsic temperature of millions of Kelvin, as is the case for the VLBI synchrotron radio emitting jet components \citep{Momjian18}. A thermal emission scenario is also in agreement with the radio--FIR spectral energy analysis presented in \cite{Rojas-Ruiz21}, which showed that the ALMA continuum emission is mainly ($>$\,90\%) of thermal origin. Assuming a typical intrinsic temperature of quasar host galaxies of $\sim$\,50\,K \cite[e.g.][]{Leipski14} implies that the emission is still significantly clumped within our 400\,pc synthesized beam \citep[as seen in the case of the $z\!=\!6.6$ quasar J0305--3150][]{Meyer2025}.

If we assume that the dust continuum is indeed powered by star formation, we can derive the corresponding star formation rates (SFRs): using standard assumptions, our 1$\sigma$ rms noise of 4.2\,$\mu$Jy\,beam$^{-1}$ corresponds to a total infrared (TIR) luminosity of L$_{\rm TIR} = 1.5 \times 10^{10}$\,L$_\odot$, or a  TIR--derived 1$\sigma$ star formation rate (SFR)  of 1.4\,M$_\odot$\,yr$^{-1}$ \citep{Kennicutt12}. I.e., the individual clumps seen in the ALMA continuum map have implied SFRs in the range of 5--10\,M$_\odot$\,yr$^{-1}$, with a central peak SFR of $\sim$13\,M$_\odot$\,yr$^{-1}$. For reference, the total TIR--based SFR in \radioqso\ is $110.0\,\pm\,13.0$\,M$_\odot$\,yr$^{-1}$ \citep{Rojas-Ruiz21}.

\subsection{Integrated ALMA \cii\ emission}

The integrated \cii\ map is shown in the right panel of Fig.~\ref{fig:alma}. We measure a total \cii\ line flux of 0.37 Jy\,km\,s$^{-1}$, or about one third of the total flux reported in \cite{Rojas-Ruiz21}. This implies that about two thirds of the \cii\ line is distributed over larger scales in the host galaxy (a slightly larger fraction than in the case of the continuum, Sec.~\ref{sec:cont}). This \cii\ map shows the same linear structure as the continuum map, albeit over a slightly smaller total extent (0.7$\arcsec$, 4.1 kpc). It is also less symmetric than the dust continuum emission, with excess \cii\ emission towards the South--East, compared to the North--West (albeit the respective S/N is not high enough to formally rule out that the emission peaks are different). We discuss the \cii\ kinematics in Sec.~\ref{PV}.

\begin{figure*}
\centering
\includegraphics[width=\textwidth]{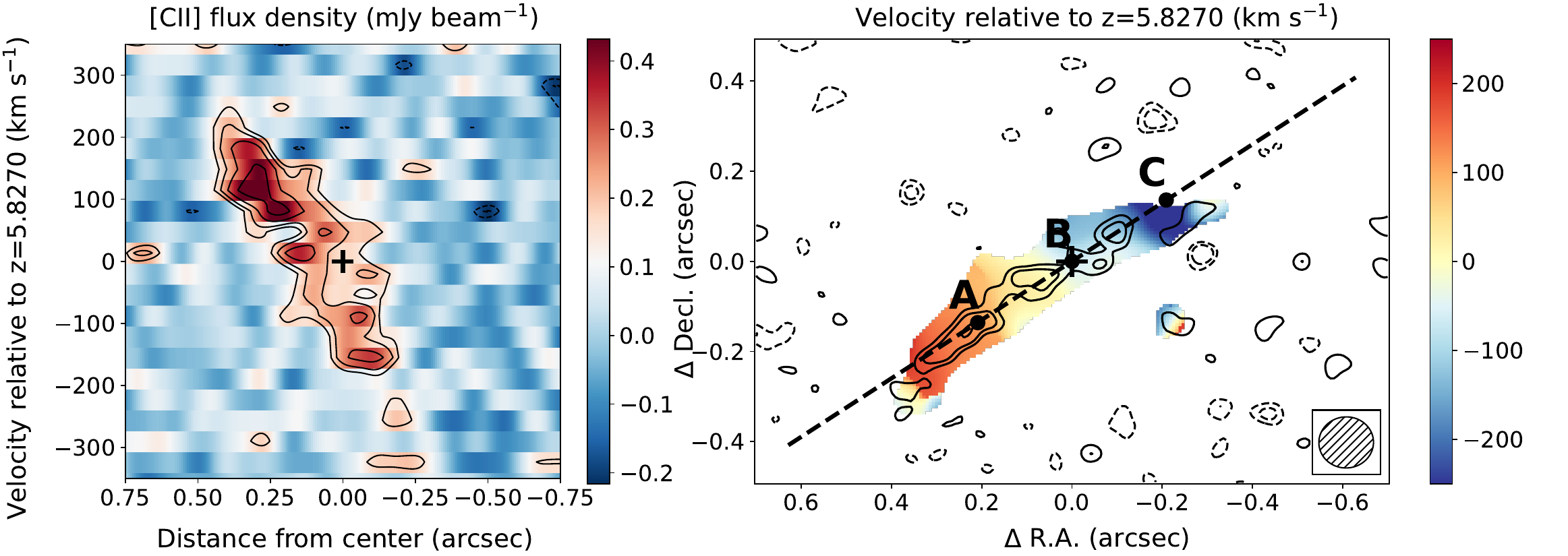}
\includegraphics[width=0.9\textwidth]{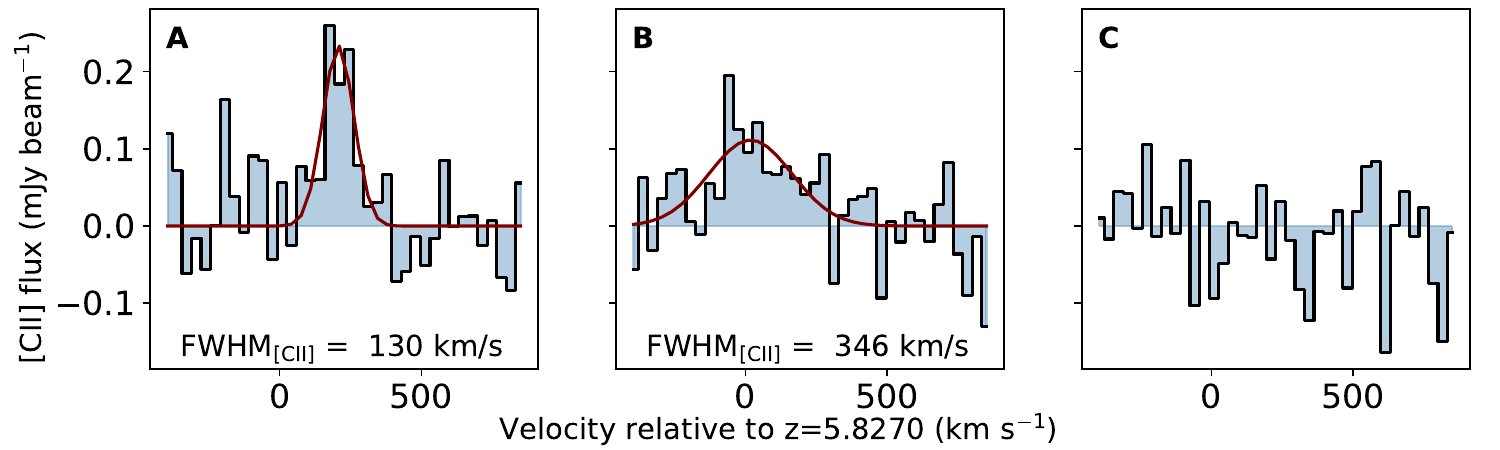}
\caption{Position--velocity diagram (\textit{top left}) of the \cii\ emission oriented along the direction of the jet as shown by the dashed line (PA=303$^\circ$) in the velocity field of the \cii\ emission line (\textit{top right}). 
In both panels, positive spatial offsets correspond to offsets towards the East. The cross indicates the position of the brightest peak in the VLBA observations \citep{Momjian18}. Contours are plotted in powers of $\sqrt{2}$, starting at $\pm$2$\sigma$ (negative contours are dashed). To increase the signal--to--noise ratio of the extended emission, we smoothed the data to a resolution of $\sim$ 0\farcs1, as shown by the updated beam in the bottom right inset of the right panel (the contours in the right panel are the integrated \cii\ flux density map using the original resolution shown in Figs.~\ref{fig:alma} and~\ref{fig:vlbi}). A smooth, linearly increase velocity gradient is apparent, reaching maximum velocities of approximately $+(-)$200\,km\,s$^{-1}$ along the Eastern (Western) jet. The bottom panels are single--beam spectra extracted from different positions along the jet (labeled A, B and C in top right panel) spaced 0\farcs25 apart. The velocity dispersion determined from a Gaussian fit to the data (red line) is given in the two bottom left panels. No successful fit was obtained for the spectrum at position `C'.
}
\label{fig:pv}
\end{figure*}

\subsection{VLBA observations}
\label{sec:vlba}

VLBA 1.54\, GHz observations of the radio jet were reported in \cite{Momjian18}, where two possible scenarios are discussed: (i) a radio core with a one-sided jet and (ii) a double--lobed symmetric object. Follow--up VLBA observations at other frequencies now greatly favor scenario (i), where the brightest  VLBI source is indeed the core of the system (Momjian et al.\ in prep.). This implies that the radio counterjet is invisible due to relativistic de--boosting effects. 

We overplot the VLBA 1.54\,GHz image on top of our ALMA observations in Fig.~\ref{fig:vlbi}. For this, we shifted the ALMA images by 24 milli--arcsecs (mas) to the West and 5\,mas to the North (less than half a synthesized ALMA beam), to align the bright central source in both the ALMA and the VLBA maps. This is justified, as the ALMA astrometry is challenging at these resolutions:  According the ALMA Technical Handbook\footnote{Chapter 10 in https://almascience.eso.org/documents-and-tools/cycle8/alma-technical-handbook}, at resolutions of $\sim$50 mas the astrometic accuracy is about $\sim$10\% of the synthesized beam if the science target is detected at S/N$>20$. In our case, the astrometric accuracy is therefore signal--to--noise--limited. As the current observations already represent a 30\,h investment with ALMA, it will be difficult to significantly increase the current signal–to–noise in the continuum before the completion of the ALMA Wideband Sensitivity Upgrade. For reference, the VLBA observations employed nodding--style phase referencing using the nearby calibrator J2327--1447, with a reported uncertainty in the position of the phase calibrator of 0.09\,mas in Right Ascension and 0.13 mas in Declination \citep{Momjian18}. Within the uncertainties, the positions of the optical quasar and the brightest ALMA and VLBA continuum source are thus in agreement.

\subsection{Morphology comparison}
\label{sec:morphology}
When overplotting all three maps (ALMA sub--millimeter dust continuum, \cii\ line emission, VLBA 1.54\,GHz radio continuum emission) in Fig.~\ref{fig:vlbi}, we find clear overall alignment between all structures in \radioqso\ along the linear structure defined by the VLBA radio jet. We stress that we see \cii\ and dust continuum emission associated with the eastern counter-jet, which is unseen in radio observations. This makes physical sense: 
the actual jet is powered on both sides of the central quasar, but only one side is seen in the radio VLBA  observations. This is because the relativistic boosting of the side of the jet pointed toward us is contrasted by a de--boosting on the opposite side, quenching observed flux. However, if the jet gives rise to enhanced dust (and gas) emission towards both sides (as argued in Sec.~\ref{sec:summary}), then this enhanced emission will be observed towards both sides in sub-millimeter emission, as no relativistic effects are at play that would boost or de--boost that emission. 

The fact that the dust continuum and \cii\ emission maxima along the jet do not show strong correspondence presumably implies that the jet encounters different physical conditions as it travels through the interstellar medium of the galaxy. We however note that there is some morphological correspondence between the jet seen in radio and dust continuum. In addition, the ALMA dust continuum image shows some symmetry in the jet and counter--jet: the three clumps on either side of the central source are spaced relatively symmetric and might indicate that the quasar activity is bursty in nature. If there was an older episode of a double radio jet, a low frequency observation might reveal symmetric large scale structure in the source \citep[e.g.,][]{Miley08}. 

\subsection{Kinematics}
\label{PV}
In the top left panel of Fig.~\ref{fig:pv}, we show a position velocity diagram of the \cii\ emission along the linear jet structure, centered on the ALMA continuum emission peak. In the top right panel of the same figure we show the \cii\ velocity field of \radioqso\ (here the dashed line indicates the orientation of the position--velocity diagram shown in the left). To create this velocity field, we have fitted Gaussian profiles to all pixels detected at $>$3\,$\sigma$ in the velocity-integrated flux density map. This approach yields more reliable results in cases of low S/N compared to a first moment image \citep[see e.g., Appendix C in ][]{Neeleman21}. A clear \cii\ velocity gradient is apparent; velocities increase approximately linearly as a function of distance from the center, from the systemic velocity of the source to $\pm$200 km\,s$^{-1}$. In creating these maps, we have revised the systemic velocity (and thus redshift) of \radioqso\ slightly, by $\sim\,-100$\,km\,s$^{-1}$, from $z=5.831$ to $z=5.8270$. This shift centers the \cii\ spectrum at the nominal position of the quasar (i.e., the point where the continuum emission is maximum). This shift is necessary because significant \cii\ emission arises from the jet, which shifts the centroid in the low--resolution data presented in \cite{Rojas-Ruiz21}.

From the \cii\ kinematics we can reach a number of important conclusions: 
most importantly, the \cii\ velocities encountered are typical for galaxy kinematics and interactions (up to a few 100\,km\,s$^{-1}$), and we do not find major velocity discontinuities at the level of thousands of km\,s$^{-1}$, which might be expected for jet-cloud interactions. (Sec.~\ref{sec:morphology}). Second, towards the center (indicated by the black cross and the spectrum at position `B'), there is evidence for enhanced velocity dispersion, albeit at low signal--to--noise. Towards the South (position A) \cii\ emission shows a velocity dispersion with a \cii\ FWHM of 130 km\,s$^{-1}$, consistent with previous dispersion measurements in other $z\gtrsim 6$ quasars based on similarly high--resolution ALMA data \citep{Venemans19, Walter22, Meyer23, Neeleman23, Meyer2025}. 

\section{Discussion and Summary}
\label{sec:summary}
We report 400\,pc resolution ALMA imaging of the \cii\ line and the  285\,GHz dust continuum in \radioqso\ at $z=5.83$, a radio--loud quasar at the end of Cosmic Reionization. The ALMA and VLBA observations reveal a remarkably narrow and linear correlation between the \cii, thermal dust, and VLBA radio synchrotron emission, with a full length of about 4 kpc and width at least ten times smaller. The gas kinematics show a simple linear velocity gradient, with velocities that are consistent with those encountered in typical galactic environments.

We interpret our findings with the following physical picture: Consider a galaxy--wide distribution of cool gas and dust in the quasar host galaxy with a large scale velocity gradient due to rotation, galaxy interactions, or infall of gas. However, this emission on galactic scales is too faint to be mapped out with our current 400\,pc resolution data. A powerful relativistic radio jet is launched by the central quasar, and rapidly passes through the host galaxy and its interstellar medium. Synchrotron emission is seen from radio jet knots, which form and fade on short time scales in regions of enhanced jet-cloud interaction \citep[on timescales of thousands of years or even less, e.g.,][]{Perucho14}. In these same regions, shocks driven by the jet compress the local clouds, heat the interstellar medium, and may even induce some clouds to form stars (`positive feedback'). In turn, this injection of turbulent energy (and possibly star formation) heats the dust, giving rise to the observed rest frame far--infrared continuum and \cii\ line emission seen by ALMA. The passage of the jet thus effectively leaves a `streak' of enhanced thermal dust continuum and \cii\ emission, which would be persistent on timescales longer than its cause (the radio jet with relativistic speeds). If this emission was associated with star formation, it would likely even persist for much longer (millions of years), than most of the radio continuum jet emission. The fact that we do not see any `streaks' of enhanced star formation in radio--quiet quasars (assuming they were radio--loud at some point in their past) would then suggest that either the time scales for this enhanced star formation are comparable to the time scales of the jet, or otherwise dynamical processes within the disk (e,g., mergers, disk rotation, etc.) quickly remove any evidence of jet--triggered enhanced star formation.

This situation resembles the well--studied radio--optical `alignment effect' in high redshift radio galaxies, in which a general alignment is seen between the kpc-scale radio jet direction and the optical starlight and Ly$\alpha$ emission on scales out to tens of kpc \citep[e.g.,][]{McCarthy90}. In most cases, this alignment is not exact, but is seen as just a broad agreement.  However, there is at least one case, the $z=3.8$ radio galaxy 4C\,41.17, where the alignment between starlight, Ly$\alpha$, [\ion{C}{1}] and CO emission, and the radio jet is remarkably close, linear, and narrow, in the inner jet on a scale of $\sim 14$ kpc \citep{Bicknell00, Nesvadba20}, similar to what is seen in \radioqso. In 4C\,41.17 these authors argue, from the close correlation between bright radio knots with optical components, as well as from the broad velocity dispersions of the \ion{C}{4} and millimeter emission lines, for jet-induced star formation, where shocks caused by the advancing radio source compress the dense clouds and enhance star formation on a timescale of order 10$^6$ years. Thus, 4C\,41.17 perhaps shows the most direct evidence for positive `radio mode feedback' in a distant galaxy, where the expanding radio source enhances the galaxy formation process through induced star formation. An alternative physical interpretation for the alignment effect was introduced by \cite{Eales92}. In this interpretation, radio hotspots are brightened when they travel through a high--density environment.

Our high-resolution ALMA imaging of \radioqso\ shows a similar, very close alignment of the axes of multiple components (in our case gas and dust) of the host galaxy with the radio jet. The \cii\ velocity dispersion along the jet in \radioqso\ is significant ($\gtrsim 100$\,km\,s$^{-1}$), hinting at kinematic evidence for jet--cloud interactions. However, there are differences between \radioqso\ and 4C\,41.17, principally relating to the fact that \radioqso\ is a broad line radio--loud quasar \citep{banados2018c}, not a radio galaxy: A VLT/X--shooter optical/NIR spectrum shows a typical high redshift quasar spectrum, including broad Ly$\alpha$ that is truncated at short wavelength due to Ly$\alpha$ resonant scattering by a significantly neutral IGM, as well as broad \ion{C}{4} emission with a line width of 4330 km s$^{-1}$, from which a black hole mass of $\sim 10^9$ M$_\odot$ has been derived (Xie et al.\ in prep.).  Further, Chandra observations show that \radioqso\ has an X-ray power-law spectrum and optical to X-ray luminosity ratio characteristic of high redshift quasars \citep{Connor21}. 

Potentially counter to the above scenario is a recent study of the integrated radio spectrum of \radioqso, which shows a clear steepening in the spectrum around an observed frequency of 30\,GHz, out to 100\,GHz (Rojas-Ruiz et al.\ subm.), from which they derive a synchrotron/inverse Compton `spectral age' for the radio source of order 500 years, depending on the magnetic field strength. This is very short compared to the timescale that would be required for jet--induced star formation to proceed (millions of years). Alternatively, the jet activity could be episodic, with previous episodes giving rise to the emission seen in the ALMA observations. In any case, the astonishingly close alignment and narrow linearity of the radio jet with the \cii\ and dust emission are hard to conceive without some fundamental relationship between the two. Reconciling the respective timescales remains paramount to understanding the nature of \radioqso. 

\begin{acknowledgments}
We thank Santiago Garcia--Burillo, Ryan Keenan, and Alessandro Lupi for useful discussions. This paper makes use of the following ALMA data: ADS/JAO.ALMA\#2018.1.00656.S and 2022.1.01587.S. ALMA is a partnership of ESO (representing its member states), NSF (USA) and NINS (Japan), together with NRC (Canada), MOST and ASIAA (Taiwan), and KASI (Republic of Korea), in cooperation with the Republic of Chile. The Joint ALMA Observatory is operated by ESO, AUI/NRAO and NAOJ. The National Radio Astronomy Observatory is a facility of the National Science Foundation operated under cooperative agreement by Associated Universities, Inc. C.M. acknowledges support from Fondecyt Iniciacion grant 11240336 and the ANID BASAL project FB210003.  RAM, MN, and FW acknowledge support from the ERC Advanced Grant 740246 (Cosmic\_Gas). RAM acknowledges support from the Swiss National Science Foundation (SNSF) through project grant 200020\_207349.

\end{acknowledgments}

\vspace{5mm}
\facilities{ALMA}

\software{
Astropy \citep{astropy2018},
CASA \citep{mcmullin2007},
Matplotlib \citep[][\url{http://www.matplotlib.org}]{hunter2007}
          }


\bibliography{references}{}
\bibliographystyle{aasjournal}



\end{document}